\documentclass[apl,twocolumn,showpacs,amsmath,letter]{revtex4}
\usepackage{graphicx}

\newcommand{\Journal}[4]{#1 \textbf{#2}, #3 (#4)}

\begin{document}

\title{Switching Current {\it vs.} Magnetoresistance in Magnetic Multilayer Nanopillars.}

\author{S. Urazhdin}
\author{Norman O. Birge}
\author{W. P. Pratt Jr.}
\author{J. Bass}
\affiliation{Department of Physics and Astronomy, Center for
Fundamental Materials Research and Center for Sensor Materials,
Michigan State University, East Lansing, MI 48824}

\pacs{73.40.-c, 75.60.Jk, 75.70.Cn}

\begin{abstract}
We study current-driven magnetization switching in nanofabricated
magnetic trilayers, varying the magnetoresistance in three
different ways. First, we insert a strongly spin-scattering layer
between the magnetic trilayer and one of the electrodes, giving
increased magnetoresistance. Second, we insert a spacer with a
short spin-diffusion length between the magnetic layers,
decreasing the magnetoresistance. Third, we vary the angle between
layer magnetizations. In all cases, we find an approximately
linear dependence between magnetoresistance and inverse switching
current. We give a qualitative explanation for the observed
behaviors, and suggest some ways in which the switching currents
may be reduced.
\end{abstract}

\maketitle

The observation~\cite{cornellorig} of
predicted~\cite{slonczewski,berger} current-driven switching in
nanofabricated magnetic multilayers (nanopillars) opened the
possibility for direct switching of the bits in magnetic memory by
local application of current, rather than by the field of external
wires. However, the present switching currents $I_s$ are too large
for high-density applications. In this paper, we describe three
new experiments that show an approximately linear dependence
between $1/I_s$ and the change of resistance $\Delta R$ upon
switching. These results should provide guidance for both theory
and engineering of current-switching devices.

First, we enhance $\Delta R$ in Py/Cu/Py/Cu
(Py=Permalloy=Ni$_{84}$Fe$_{16}$) trilayer nanopillars by
inserting $1$~nm of a strong spin-scatterer,
Fe$_{50}$Mn$_{50}$~\cite{sdlength} between the trilayer and the
top electrode. Second, we insert a $t_{CuPt}$ thick
Cu$_{96}$Pt$_6$ layer between the Py layers. The short
spin-diffusion length in Cu$_{94}$Pt$_6$ decreases $\Delta R$.
Third, we study $\Delta R$ and the switching currents $I_s$ as a
function of the angle between the magnetizations of the two
ferromagnetic layers in Py/Cu/Py nanopillars.

\begin{figure}
\includegraphics[scale=0.4]{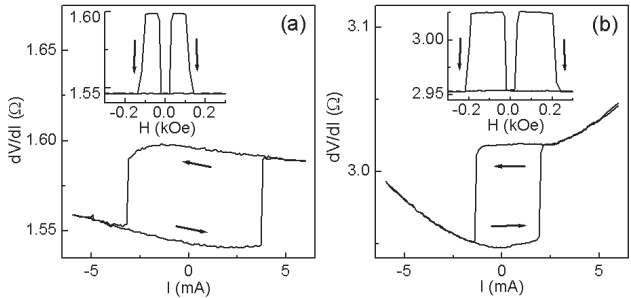}
\caption{\label{fig1} (a) $dV/dI$ {\it vs.} $I$ for a sample of
type 1 (as defined in the text) at $H=0$. Inset: $dV/dI$ {\it vs.}
$H$ at $I=0$. (b) Same as (a), for a sample of type 2.}
\end{figure}

Our samples were made with a multistep process described
elsewhere~\cite{myapl}. Below, all thicknesses are in nm. The
basic samples had structure
Cu(80)/F$_1$=Py(30)/N(15)/F$_2$=Py(6)/Cu(2)/Au(150). The bottom
Cu(80)/Py(30) layers were extended leads, N, F$_2$ and Cu(2) were
patterned into an elongated shape with dimensions $\approx
130\times 70$~nm, and Au(150) was the top lead. Leaving F$_1$
extended minimizes the effect of dipolar coupling on the
current-driven switching~\cite{cornellapl}. N was
Cu(13.5-d)/Cu$_{94}$Pt$_6$(d)/Cu(1.5), with d=0, 0, 4, 8, 12 in
sample types 1 through 5, respectively.  In sample type 2, the
Cu(2) layer was replaced with a Cu(2)/Fe$_{50}$Mn$_{50}$(1)/Cu(2)
sandwich. We measured $dV/dI$ at room temperature (295~K) with
four-probes and lock-in detection, adding an ac current of
amplitude 20--40~$\mu$A at 8~kHz to the dc current $I$. At least 7
samples of each type were tested. Typical sample resistances were
1 to 3~$\Omega$. Variations in resistances are attributed to
scatter in both nanopillar sizes and contact resistances to the
electrodes. Positive current flows from the extended to the
patterned Py layer. $H$ is in the film plane and (except for the
angular dependence studies) along the nanopillar easy axis.

\begin{figure}
\includegraphics[scale=0.4]{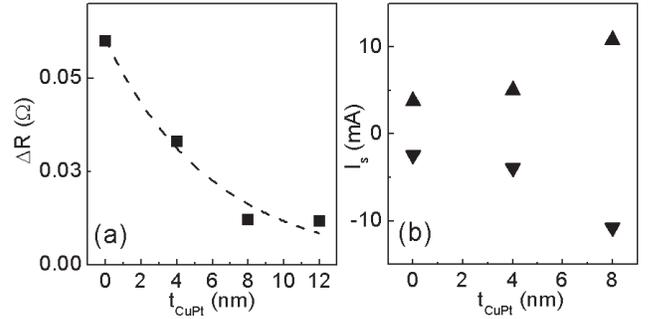}
\caption{\label{fig2} (a) Variation of $\Delta$R with $t_{CuPt}$.
Dashed line is a fit with $\Delta R=\Delta
R_0exp[-t_{CuPt}/l_{sf}^{CuPt}]$, $\Delta R_0=0.06\pm
0.004\Omega$, $l_{sf}^{CuPt}=6.1\pm 0.8$~nm. (b) $I_s^{P\to AP}$
(upward triangles) and $I_s^{AP\to P}$ (downward triangles) {\it
vs.} $t_{CuPt}$.}
\end{figure}

Fig.~\ref{fig1} compares typical results for a sample of type 1
(Fig.~\ref{fig1}(a)) {\it vs.} a sample of type 2,
(Fig.~\ref{fig1}(b)). In both cases, negative current $I$ leads to
transition from the antiparallel (AP) state with high resistance
$R_{AP}$ to the parallel (P) state with low resistance $R_P$ at
$I=I_s^{AP\to P}$. A reverse transition occurs at positive
$I=I_s^{P\to AP}$. In the $H$ sweep at $I=0$ (insets
Fig.~\ref{fig1}(a,b)), the extended $F_1$ layer reverses at
$H\approx 20$~Oe, and $F_2$ reverses at the field $H_s\approx
100-200$~Oe, determined by the shape anisotropy of F$_2$. The
average $H_s$ in samples of type~1 and type~2 were similar, insets
of Fig.~\ref{fig1} only illustrate scatter among samples. There
was no systematic correlation between $I_s$ and $H_s$. Since the
high resistivity $\rho\approx 100\mu\Omega$cm of
Fe$_{50}$Mn$_{50}$ contributes only $\approx0.25$~$\Omega$ to the
resistance of samples of type 2, the contact resistance in
Fig.~\ref{fig1}(b) must be $\approx 1\Omega$ larger than that in
Fig.~\ref{fig1}(a). For 14 samples of type 1, $\Delta R\equiv
R_{AP}-R_{P}=0.060\pm 0.002\Omega$, $I_s^{AP\to P}=-2.45\pm
0.2$~mA, and $I_s^{P\to AP}=3.8\pm 0.2$~mA.  For 12 samples of
type 2, $\Delta R=0.085\pm 0.012\Omega$, $I_s^{AP\to P}=-1.5\pm
0.2$~mA, and $I_s^{P\to AP}=1.85\pm 0.2$~mA. For uncertainties, we
give twice the standard deviations of the mean. The main result of
this experiment is the higher $\Delta R$, and lower $I_s$, in the
nanopillars with the inserted Fe$_{50}$Mn$_{50}$(1) layer.

Fig.~\ref{fig2} shows data for sample types 1, 3, 4, 5.
Fig.~\ref{fig2}(a) ($\Delta$R($t_{CuPt})$) gives a spin-diffusion
length of $6.1\pm 0.8$~nm in Cu$_{94}$Pt$_6$ at 295~K, shorter
than $\approx10$~nm at 4.2~K~\cite{sdlength}. Fig.~\ref{fig2}(b)
shows that both $I_s^{P\to AP}$ and $|I_s^{AP\to P}|$ increase
with increasing $t_{CuPt}$. Interestingly, the ratio $I_s^{P\to
AP}/|I_s^{AP\to P}|$ decreases from $\approx 1.5$ at $t_{CuPt}=0$
to $\approx 1.0$ for $t_{CuPt}=8$. This decrease with increase of
spin-flipping within the N-layer is opposite to that reported
in~\cite{cornellquant} for a similar measurement with varied
thickness of N=Cu, and is inconsistent with the explanation
proposed there. All 8 samples of type 5 showed hysteretic
field-driven switching, similar to other sample types. However,
none showed reproducible hysteretic current-switching. Such a
qualitative change at sufficiently large $t_{CuPt}$ is expected.
When the Py layers are nearly decoupled due to spin-flip
scattering in a thick Cu$_{94}$Pt$_6$ layer, the effect of current
becomes independent of the mutual orientations of these layers.
This effect is similar to that for a single magnetic layer, and
cannot lead to hysteretic switching between P and AP states.

\begin{figure}
\includegraphics[scale=0.4]{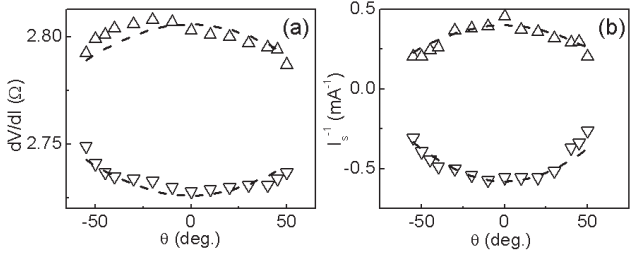}
\caption{\label{fig3} (a) Quasi-parallel (P) and
quasi-antiparallel (AP) state resistances $R_P$ (upward triangles)
and $R_{AP}$ (downward triangles) {\it vs.} $\theta$. Dashed
curves: fits with $R_{P,AP}=R_0\mp A\cos\theta$, $R_0=2.77\Omega$,
$A=0.04\Omega$. (b) $1/I_s^{P\to AP}$ (upward triangles) and
$1/I_s^{AP\to P}$ (downward triangles) {\it vs.} $\theta$. Dashed
curves: fits with $1/I_s^{P\to AP,AP\to P}=K_{P,AP}\cos\theta$,
$K_P=0.40$~mA$^{-1}$, $K_{AP}=-0.58$~mA$^{-1}$.}
\end{figure}

Fig.~\ref{fig3} shows the results for varied non-collinear
orientations of magnetic layers in sample type 1. Before each
measurement, a pulse of $H=60$~Oe at the desired in-plane angle
$\theta$ was applied to rotate the magnetization M$_1$ of F$_1$
parallel to $H$, then the current-switching was measured at
$H=10$~Oe, needed to fix M$_1$. The data in Fig.~\ref{fig3}
confirm results reported in~\cite{mancoff}, but with a larger
magnetoresistance $\Delta R/R$.

\begin{figure}
\includegraphics[scale=0.4]{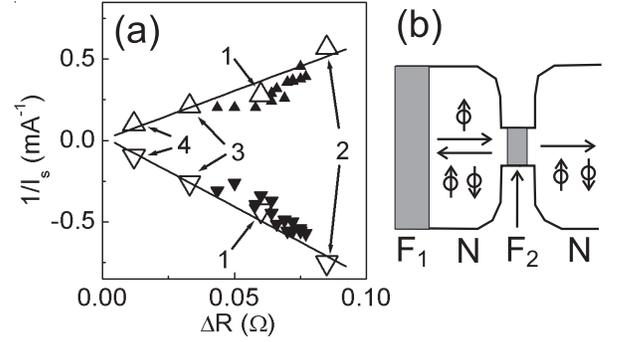}
\caption{\label{fig4} (a) Dependence of $1/I^{P\to AP}_s$ (upward
triangles) and $1/I^{AP\to P}_s$ (downward triangles) on $\Delta
R$. Open symbols: sample types 1 through 4, as labeled. Solid
symbols: variations with angle between the magnetizations in a
sample of type 1. Solid lines: best linear fits of data, excluding
the angular dependence. The ordinate intercepts are zero within
the uncertainty of the fits, (b) Schematic of electron scattering
in nanopillar, as discussed in the text.}
\end{figure}

In Fig.~\ref{fig4}(a) we collect together the data of
Figs.~\ref{fig1}-\ref{fig3} in a plot of average values of $1/I_s$
{\it vs.} average $\Delta$R. The variations among different
samples lead to uncertainties of the average values, close to the
symbol sizes in Fig.~\ref{fig4}(a). The overall agreement of the
data for three different types of measurements suggests a general
inverse relationship between $I_s$ and $\Delta R$, independent of
the particular way in which $\Delta R$ was varied. The switching
is determined by the current density, so both $1/I_s$ and $\Delta
R$ are inversely proportional to the nanopillar areas; their
variation only leads to scaling along the approximately linear
dependence in Fig.~\ref{fig4}(a).

To qualitatively describe the inverse relationship in
Fig.~\ref{fig4}(a), we use the simplest plausible ballistic model,
in which the electrons polarized by F$_1$ are scattered in F$_2$,
generating magnetic excitations. Fig.~\ref{fig4}(b) shows a
cartoon of this model, where a spin-up electron coming from F$_1$
is either transmitted or reflected by F$_2$. In either case, it
can flip its spin, exciting (or de-exciting) the F$_2$ layer. We
introduce a parameter $p$, describing the polarization of current
if F$_2$ is removed. We define the sign of $p$ with respect to the
direction of magnetization M$_2$ of F$_2$. For Py (sample type 1),
we expect $p\approx 0.45-0.6$~\cite{pratt,nadgorny}. When F$_1$ is
absent, $p=0$. The current polarization in the diffusive transport
model~\cite{kovalev} is different; it depends both on F$_1$ and
F$_2$, and does not disappear when F$_1$ is absent.

In our model, $\Delta R$ is determined by the spin-dependent
resistance of the interfaces and bulk of F$_2$ and is proportional
to $p$. We can interpret the variations of $\Delta R$ in terms of
a change in polarization $p$. When Cu$_{94}$Pt$_6$ is inserted
between F$_1$ and F$_2$, spin-flip scattering in this layer
decreases $p$ according to
$p(t_{CuPt})=p(0)exp[-t_{CuPt}/l_{sf}^{CuPt}]$, consistent with
Fig.~\ref{fig2}(a). The angular dependence of Fig.~\ref{fig3}(a)
can be understood similarly in terms of the projection of spin
current onto the direction of the magnetization of F$_2$, giving
$p(\theta)=p(0)cos(\theta)$. Finally, the Fe$_{50}$Mn$_{50}$
inserts outside F$_2$ have very short spin-diffusion length.
Although the resulting increase of MR involves spin-diffusion
outside the magnetic trilayer, and cannot by described by our
ballistic model, it is also reasonable to approximate the effect
of the Fe$_{50}$Mn$_{50}$ inserts as an increase of $p$.

The current-driven switching is also expected to be determined by
$p$. Electrons with spin opposite to $M_2$ can generate magnetic
excitations when they flip their spins, while electrons with spins
along the magnetization can absorb the excitation when they
spin-flip, as follows from the conservation of angular momentum
along M$_2$~\cite{berger}. Thus, we may expect the rate of
magnetic excitation by current to be given by the difference
between the spin-down and spin-up electron currents, i.e.
approximately proportional to $p\cdot I$.  $p\cdot I_s$ is then
determined by the level of magnetic excitation, needed for the
magnetization switching, i.e. $1/I_s\propto p$. The data and
linear fits (solid lines) in Fig.~\ref{fig4}(a) are consistent
with this analysis. Our data are also generally consistent with
the more quantitative analyses of current-driven switching based
on the popular spin-torque model~\cite{slonczewski,kovalev}, and
the recently proposed effective temperature model~\cite{mytheory}.
These models differ from each other in details, which need further
experimental testing.

Our data, and the simple model, suggest that one might reduce
$I_s$ by using a more highly polarizing ferromagnet for F$_1$, or
by being more clever in designing the layers 'outside' the
F$_1$/N/F$_2$ trilayer. Independent evidence that the
current-driven switching is determined by the N/F$_2$
interfaces~\cite{cornellquant} suggests that modifying those
interfaces (e.g. by varying their roughness or local composition)
should be worth exploring. At room temperature, the current-driven
switching is thermally activated~\cite{cornelltemp, myprl}. An
obvious way of decreasing the switching current is then to lower
the switching barrier. But a smaller switching barrier also leads
to thermal activation at room temperature without applied current,
thus reducing the effectiveness of the nanopillars for information
storage.

To summarize, we measured the changes in resistance upon
switching, $\Delta R$, and the switching currents, $I_s$, in
Permalloy (Py)-based trilayer nanopillars with: a) strong spin
flipping between the nanopillar and one of the leads, b)
spin-flipping in the spacer between the Py layers, c) varying
angle between the magnetizations of the Py layers. We find a
linear relation between $I_s^{-1}$ and $\Delta R$. We describe the
data in terms of a qualitative ballistic model.

This work was supported by the MSU CFMR, CSM, the MSU Keck
Microfabrication facility, the NSF through Grants DMR 02-02476,
98-09688, and NSF-EU 00-98803, and Seagate Technology.


\begin{thebibliography}{99}
\bibitem{cornellorig} J.A. Katine, F.J. Albert, R.A. Buhrman, E.B. Myers and D.C. Ralph, \Journal{Phys. Rev.
Lett.}{84}{3149}{2000}, and references in J. Bass, S. Urazhdin,
N.O.Birge, and W.P. Pratt Jr., Phys. Stat. Sol. (in press).
\bibitem{slonczewski} J. Slonczewski, \Journal{J. Magn. Magn. Mater.}{159}{L1}{1996}.
\bibitem{berger} L. Berger, \Journal{Phys. Rev.}{B 54}{9353}{1996}.
\bibitem{sdlength} W. Park, D.V. Baxter, S. Steenwyk, I. Moraru,
W.P. Pratt Jr., and J. Bass, \Journal{Phys. Rev.}{B
62}{1178}{2000}.
\bibitem{myapl} S. Urazhdin, H. Kurt, W.P. Pratt Jr., and J. Bass, \Journal{Appl. Phys.
Lett.}{83}{114}{2003}.
\bibitem{cornellapl} F. J. Albert, J.A. Katine, R.A. Buhrman, and D.C. Ralph, \Journal{Appl. Phys. Lett.}{77}{3809}{2000}.
\bibitem{cornellquant} F. J. Albert, N.C. Emley, E.B. Myers, D.C. Ralph, and R.A. Buhrman, \Journal{Phys. Rev. Lett.}{89}{226802}{2002}.
\bibitem{mancoff} F.B. Mancoff, R.W. Dave, N.D. Rizzo, T.C.
Eschrich, B.N. Engel, and S. Tehrani, \Journal{Appl. Phys.
Lett.}{83}{1596}{2003}.
\bibitem{pratt} W. P. Pratt Jr., S.D. Steenwyk, S.Y. Chiang, A.C. Scaefer, R. Loloee, and J. Bass,
\Journal{IEEE Trans. Magn.}{33}{3505}{1997}.
\bibitem{nadgorny} B. Nadgorny, R.J. Soulen Jr., M.S. Osofsky, I.I. Mazin, G. Laparade, R.J.M. van de Veerdonk,
A.A. Smits, S.F. Cheng, E.F. Skelton, and S.B. Qadri,
\Journal{Phys. Rev.}{B 61}{R3788}{2000}.
\bibitem{kovalev} A.A. Kovalev, A. Brataas, and G.E.W. Bauer,
\Journal{Phys. Rev}{B 66}{224424}{2002}.
\bibitem{mytheory} S. Urazhdin, cond-mat/0308320.
\bibitem{cornelltemp} E. B. Myers, F.J. Albert, J.C. Sankey, E. Bonet,
R.A. Buhrman, and D.C. Ralph, \Journal{Phys. Rev.
Lett.}{89}{196801}{2002}.
\bibitem{myprl} S. Urazhdin, N.O. Birge, W.P. Pratt Jr., and J. Bass, Phys. Rev. Lett. (in press), cond-mat/0303149.
%%\bibitem{myunpublished} S. Urazhdin, N.O. Birge, W.P. Pratt Jr., and J. Bass, unpublished.
%%\bibitem{cornellscience} E.B. Myers, D.C. Ralph, J.A. Katine, R.N.Louie, R.A. Buhrman, \Journal{Science}{285}{867}{1999}.
%%\bibitem{chien} Y. Ji, C.L. Chien, and M.D. Stiles, \Journal{Phys. Rev. Lett.}{90}{106601}{2003}.
%%\bibitem{zhang2} Z. Li and S. Zhang,  cond-mat/0302339.
%%\bibitem{effecttemp}R.H. Koch, J.A. Katine, and J.Z. Sun, preprint.
%%\bibitem{circuit} A. Brataas, Yu.V. Nazarov, and G.E.W. Bauer, \Journal{Phys. Rev. Lett.}{84}{2481}{2000}.
\end{thebibliography}
\end{document}